\documentclass[letterpaper, 10 pt, conference]{ieeeconf}  

\IEEEoverridecommandlockouts                              

\usepackage{graphics,graphicx} 
\usepackage{amsmath,amssymb}
\usepackage{booktabs}
\usepackage{cite}
\usepackage{subcaption}
\usepackage[font=footnotesize]{caption}
\usepackage{romannum}
\usepackage{comment}
\usepackage{algorithm2e}
\usepackage{chemarrow}
\usepackage[none]{hyphenat}

\title{\LARGE \bf
Stochastic multi-step cell size homeostasis model for cycling human cells}
\author{Sayeh Rezaee$^{1}$, Cesar Nieto$^{2}$, and Abhyudai Singh$^{3}$
\thanks{$^{1},^{2}$ Department of Electrical and Computer Engineering,
              University of Delaware, Newark, DE, USA
        {\tt\small \{sayehr, cnieto\}@udel.edu}}%
\thanks{$^{3}$ Department of Electrical and Computer Engineering, Biomedical Engineering, Mathematical Sciences, Center of Bioinformatic and Computational Biology, University of Delaware, Newark, DE, USA
        {\tt\small\{absingh\}@udel.edu}}%
}

\begin{document}

\maketitle
\thispagestyle{empty}
\pagestyle{empty}

\begin{abstract}

Measurements of cell size dynamics have established the adder principle as a robust mechanism of cell size homeostasis.
In this framework, cells add a nearly constant amount of size during each cell cycle, independent of their size at birth.
Theoretical studies have shown that the adder principle can be achieved when cell-cycle progression is coupled to cell size.
Here, we extend this framework by considering a general growth law modeled as a Hill-type function of cell size. This assumption introduces growth saturation to the model, such that very large cells grow approximately linearly rather than exponentially.
Additionally, to capture the sequential nature of division, we implement a stochastic multi-step adder model in which cells progress through internal regulatory stages before dividing.
From this model, we derive exact analytical expressions for the moments of cell size distributions.
Our results show that stronger growth saturation increases the mean cell size in steady state, while slightly reducing fluctuations compared to exponential growth. Importantly, despite these changes, the adder property is preserved.
This emphasizes that the reduction in size variability is a consequence of~the growth law rather than simple scaling with mean size.
Finally, we analyze stochastic clonal proliferation and find that growth saturation influences both single-cell size statistics and variability across populations.
Our results provide a generalized framework for connecting multi-step adder mechanisms with proliferation dynamics, extending size control theory beyond exponential growth.

\end{abstract}

\section{Introduction}

Cell size is a fundamental physiological property that influences several cellular processes, such as metabolism, gene expression, and proliferation. To maintain proper function, cells actively control their size, ensuring stability across generations \cite{lin2018homeostasis,schwabe2011origins,rhind2021cell,ji2023implications,ferrezuelo2012critical,vargas2016conditions, nieto2023coupling}.
Notably, cells across diverse organisms including bacteria, yeast, and animal cells exhibit substantial intrinsic random variability (noise), reflecting the inherently stochastic nature of intracellular processes \cite{hatton2023human, puliafito2017cell, cadart2018size, kiriakopulos2025lncrna, xie2025g1, fu2025cells,miller2023fission}.
Misregulation of cell size control results in increased size heterogeneity, which is a hallmark of~malignancy in many diseases, such as different cancer types~\cite{ginzberg2015being,li2015cancer}. Although having a noisy size, healthy cells maintain stable size distributions across generations, and despite extensive experimental studies, the mechanisms underlying size homeostasis remain incompletely understood. 

\begin{figure}[t]
	\centering
\includegraphics[width=0.8
\linewidth]{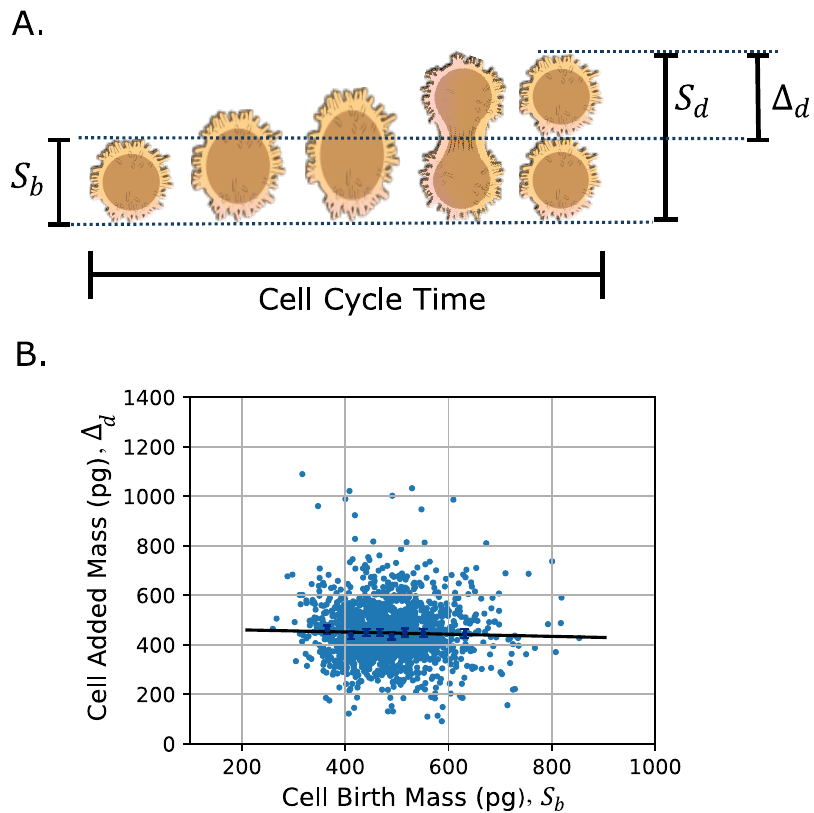}
\caption{\textbf{Visualization of the cell cycle variables and an example of adder division strategy.} 
\textbf{A)} A cell begins the cycle at its birth size $S_b$ and grows until it reaches its division size $S_d$. The increase in size during this cycle is denoted as $\Delta_d$, and the duration of the process is the cell-cycle time. Note that cell size may refer to different cellular properties, such as area or mass, depending on the context. The cell icon here represents a human B cell as an example and was adapted from NIH BioArt (public domain). 
\textbf{B)} Example dataset from \cite{liu2024cell} showing the relationship between added mass and birth mass in HeLa human cells. Each dot represents a single cell measurement. The almost zero correlation indicates that the division strategy in this dataset follows an adder mechanism, where added mass is independent of birth~mass.} \label{fig:Bio}
\end{figure}

To maintain size homeostasis, cells must coordinate growth with division, regulated by complex processes such as protein synthesis/degradation and cell cycle checkpoints~\cite{liu2024cell,liu2025oversized,willis2020limits,tan2021cell}. These diverse mechanisms can be classified into three broad division strategies: adder, where cells add a constant size each cycle, independent of their birth size; sizer, where cells divide upon reaching a critical size; and timer, where division occurs after a fixed time \cite{miotto2024size,pandey2020exponential,taheri2015cell,jones2023first,nieto2025joint}. Phenomenologically, these strategies are distinguished by the slope between added size and size at birth: 0 for an adder, –1 for a sizer, and 1 for a timer \cite{iyer2014scaling,vargas2018cell,ghusinga2016mechanistic,nieto2025dynamical} (See Fig. \ref{fig:Bio}). 

We aim to develop a quantitative framework for modeling cellular growth and division.
Classical models of cell growth often assume exponential or linear dynamics coupled with size-dependent division events. However, biological evidence shows that cellular growth does not scale indefinitely with size. In bacteria, given their small cell size and rapid nutrient uptake, exponential growth is a good assumption. However, in animal cells, metabolic constraints and the complexity of biochemical reactions impose a saturating dynamics \cite{vargas2020modeling, belliveau2021fundamental,bren2013last,ginzberg2018cell}.
These physiological constraints naturally motivate the use of nonlinear growth laws, such as Hill-type functions, which increase with size at small scales but slow down at larger cell sizes.

In parallel, studies of cell division control have revealed the adder strategy as a central organizing principle across many organisms \cite{sauls2016adder,campos2014constant}.
In a previous contribution, we showed that this division strategy can be mechanistically explained if cell-cycle regulators are produced in proportion to the growth law (defined as the time derivative of size) and division is triggered once these regulators reach a threshold \cite{nieto2024generalized}, thereby proposing a generalized adder framework.  
In this article, we extend this model to cells exhibiting non-exponential growth. This extension is motivated by recent experiments in mammalian cells showing that, in addition to growing with a size-saturating growth law, cells divide according to an adder mechanism. These concepts can also be integrated in mechanistic frameworks where cycle progression involves multiple regulatory steps and checkpoints that buffer random fluctuations and reduce cell size variability \cite{si2019mechanistic,nieto2024mechanisms}.

To formalize these assumptions, we employ stochastic hybrid systems (SHS), providing a quantitative framework for studying coupled continuous growth and discrete stochastic events~\cite{totis2020population,nieto2020unification,vargas2018elucidating,rezaee2025inferring}.
In this approach, cell size increases continuously, while regulatory transition steps to division occur as random, discrete events. In a recent work, we showed that extending SHS models to include multiple intermediate steps between successive divisions captures additional biological realism and leads to further reductions in size variability~\cite{nieto2024moments}. 
Building on this, we present a unified framework for understanding how multi-step adder mechanisms and biological constraints on growth jointly ensure robust size~homeostasis.

We first introduce the model of cell size dynamics and solve the size statistics (statistical moments) for a single-step system.
This analysis is subsequently extended to a more biologically relevant multi-step regulation, leading to analytical expressions for the corresponding statistics.
We finally examine the statistics of clonal proliferation, linking single-cell size regulation to population-level variability exploring how different levels of non-linearity can affect the cell proliferation dynamics.


\textit{Notation}: Angular brackets $\langle \; \rangle$ denote expectations of random variables and processes. A bar indicates steady-state,~i.e., $\overline{\langle \; \rangle}$ is the expected value when $t \to \infty$. $\log$ represents the logarithm in base $e$.

\section{SHS Framework for Cell Size Dynamics with Single-step Division}

Let $s(t)$ denote the cell size at time $t$. The dynamics of cell size growth can be expressed as:
\begin{equation}
\frac{ds}{dt} = f(s),
\end{equation}
where $f(s)>0$ is an arbitrary continuous function known as the growth law. 
Each cell cycle begins with an initial random size 
$s_b$, referred to as the size at birth. As the cell grows, it reaches a random division size 
$s_d$, at which point it divides into two daughter cells, each inheriting approximately half of the mother cell’s size. In the single-cell analysis presented here, we follow only one of these descendant cells across successive generations.
The added size, denoted by 
$\Delta\geq0$, quantifies the increase in cell size from birth to division within a given cycle. It starts at zero at birth and increases to 
$\Delta_d$, the added size at division, immediately before cell division. By definition, the added size satisfies 
$\Delta=s-s_b$, where the birth size 
$s_b$ can vary from cycle to cycle~(Fig.~\ref{fig:Bio}A).

Cell division is modeled as a stochastic jump process in which the probability that a cell divides within an infinitesimal time interval $(t, t+dt]$ is given by $h(s) dt$. Additionally, $h(s)$ denotes the division rate and explicitly depends on the cell size $s$. Together with the growth law $f(s)$,
these two size-dependent functions define the rules governing cell-cycle progression and size control.
A commonly studied choice for the growth law is exponential growth, $f(s) = \mu s$. In this case, if division is also modeled as a function of size, $h(s) \propto s$, the system reduces to the classical adder framework explored in earlier studies, in which both growth and division scale proportionally with cell size and are effectively unconstrained.

Here, to capture the effect of biochemical constraints on growth, we consider a Hill-type size-saturating growth law~function: 
\begin{equation}\label{eq:grrate}
    \frac{ds}{dt} = \frac{\mu s}{1 + \alpha s},
\end{equation}
where $\alpha>0$ is a saturation coefficient controlling the degree of growth saturation and $\mu>0$ is the growth constant. The division rate is modeled as:
\begin{equation}\label{eq:divrate}
    h(s)=\frac{k s}{1 + \alpha s},
\end{equation}
where $k>0$ denotes the division rate constant. This rate reflects a division probability that increases with cell size and saturates for very large cells, capturing the biological reality of finite biosynthetic capacity.
A division rate proportional to the growth law, as in Eq.~(3), arises naturally when division is coupled to the synthesis of a regulator. When produced at a rate proportional to growth, this regulator preserves the adder property~\cite{nieto2024generalized}.
Notably, \eqref{eq:divrate} bridges two distinct regimes: in the limit $\alpha s \ll 1$ (the exponential adder regime), the Hill functions for the growth law and division rate scale linearly with size as $\mu s$ and $ks$, respectively. Conversely, in the limit $\alpha s \gg 1$ (the saturation regime), these functions converge to the constant values $\frac{\mu}{\alpha}$ and $\frac{k}{\alpha}$.

\begin{figure}[h]
	\centering
\includegraphics[width=0.45
\linewidth]{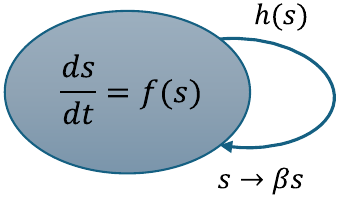}
\caption{ \textbf{Cell growth and division captured using SHS modeling framework for a single-step division system.} The size grows continuously with a growth law $f(s)$, and division is represented as a random jump occurring at rate $h(s)$. The cell size undergoes a discrete jump, modeled as $s\rightarrow \beta s$, where $\beta$ is a random variable with mean 
$\langle \beta \rangle$.
}
	\label{fig:single}
\end{figure}

Upon division, the mother cell divides into two daughter cells, and we keep track of one of them. Thus, the size reset map is given by:
\begin{equation}
s \rightarrow \beta s,
\end{equation}
where $\beta \in (0,1)$ is a random partitioning coefficient. To capture variability in division asymmetry, we assume $\beta$ follows a distribution (usually beta) with mean $\langle \beta \rangle$ and second order moment of $\langle \beta^2 \rangle$. This formulation integrates deterministic, size-dependent growth with stochastic variability arising from both division timing and partitioning (See Fig. \ref{fig:single}).

\section{Derivation of statistical moments in single-step model}
Building on this model, our goal is to characterize the statistical properties of cell size and investigate how they are affected by saturation in the growth law. To this end, we adopt a general framework for describing the moment dynamics of the proposed stochastic hybrid system. The time dynamics of the expected value of a stochastic hybrid variable is given by \cite{HS05}:
\begin{equation} \label{eq:dynkin}
    \dfrac{d\left\langle F(s) \right\rangle}{dt}= \left\langle \dfrac{ds}{dt}\cdot \dfrac{dF}{ds} + h(s)  \left( {F(\beta s)-F(s) }\right) \right\rangle,
\end{equation}
where $F(s)$ is an arbitrary function of the cell size $s$, chosen to extract the corresponding statistical moments (e.g., $F(s)=s$ yields the mean size, and $F(s)=s^2$ yields the second-order moment).
In addition, the term $F(\beta s)-F(s)$ captures the discrete jump in $F(s)$ caused by division.

We begin our analysis by applying this framework to the single-step SHS model introduced earlier. Using \eqref{eq:dynkin}, we derive the following moment dynamics:
\begin{subequations}\label{eq:sseqs}
    \begin{align}
    &\dfrac{d\left\langle \log s\right\rangle}{dt}=\mu \left\langle \dfrac{1}{1+\alpha s} \right\rangle + k\left\langle \log \beta \right\rangle \left\langle \dfrac{s}{1+\alpha s} \right\rangle,
\\
    &\dfrac{d\left\langle s \right\rangle}{dt}=\mu \left\langle \dfrac{s}{1+\alpha s} \right\rangle + k\left(\left\langle \beta \right\rangle-1\right) \left\langle \dfrac{s^2}{1+\alpha s} \right\rangle, 
\\
    &\dfrac{d\left\langle s^2 \right\rangle}{dt}=2\mu \left\langle \dfrac{s^2}{1+\alpha s} \right\rangle + k\left(\left\langle \beta^2 \right\rangle-1\right) \left\langle \dfrac{s^3}{1+\alpha s} \right\rangle.
\end{align}
\end{subequations}
By setting the left-hand sides of the above equations to zero, we solve for the corresponding variables in the steady-state limit ($t\rightarrow\infty$) and obtain closed-form expressions for the first- and second-order moments of cell size in steady-state (see notation at the end of the introduction section). To achieve this without introducing any approximations, we use the following algebraic manipulation:
\begin{equation}\label{eq:trick}
\scriptstyle
     \overline{ \left\langle \dfrac{s^n}{1+\alpha s} \right\rangle }=\dfrac{1}{\alpha} \left(\overline{ \left\langle s^{n-1}\right\rangle}-\overline{ \left\langle\dfrac{s^{n-1}}{1+\alpha s}\right\rangle } \right) .
\end{equation}
The replacement of \eqref{eq:trick} into \eqref{eq:sseqs} in steady-state, results in the first- and second-order moments of the steady-state cell size given, respectively, by:
\begin{equation}
     \overline{ \left\langle s \right\rangle } =   \dfrac{\mu \left( \mu \alpha - k  \left( \langle \beta  \rangle -1\right) \right)}{k \left( \left\langle \beta \right\rangle-1 \right) \left(k \left\langle \log \beta \right\rangle - \alpha \mu \right)},
\end{equation}

\begin{equation}
    \overline{ \left\langle s^2 \right\rangle} =   \dfrac{\mu^2 \left(  k \left( \langle \beta^2 \rangle -1 \right) -2\mu \alpha \right)}{k^2 \left( \left\langle \beta \right\rangle+1 \right) \left\langle (\beta-1)^2 \right\rangle\left(k \left\langle \log \beta \right\rangle - \alpha \mu \right)}.
\end{equation}
To quantify the noise in cell size, we use the definition of coefficient of variation:
\begin{equation} \label{eq:cv}
    CV^2_s:=\dfrac{\overline{\left\langle s^2 \right\rangle}-\overline{\left\langle s \right\rangle} ^2}{\overline{\left\langle s \right\rangle} ^2}.
\end{equation}

Next, we expand our analysis to a multi-step system comprising  $M$ steps before reaching the division. We aim to study how changes in saturation coefficient and step sizes affect the noise in cell size.

\section{Extending the Cell Size Model to Multi-step Division}

\begin{figure*}[t]
	\centering
\includegraphics[width=0.9
\linewidth]{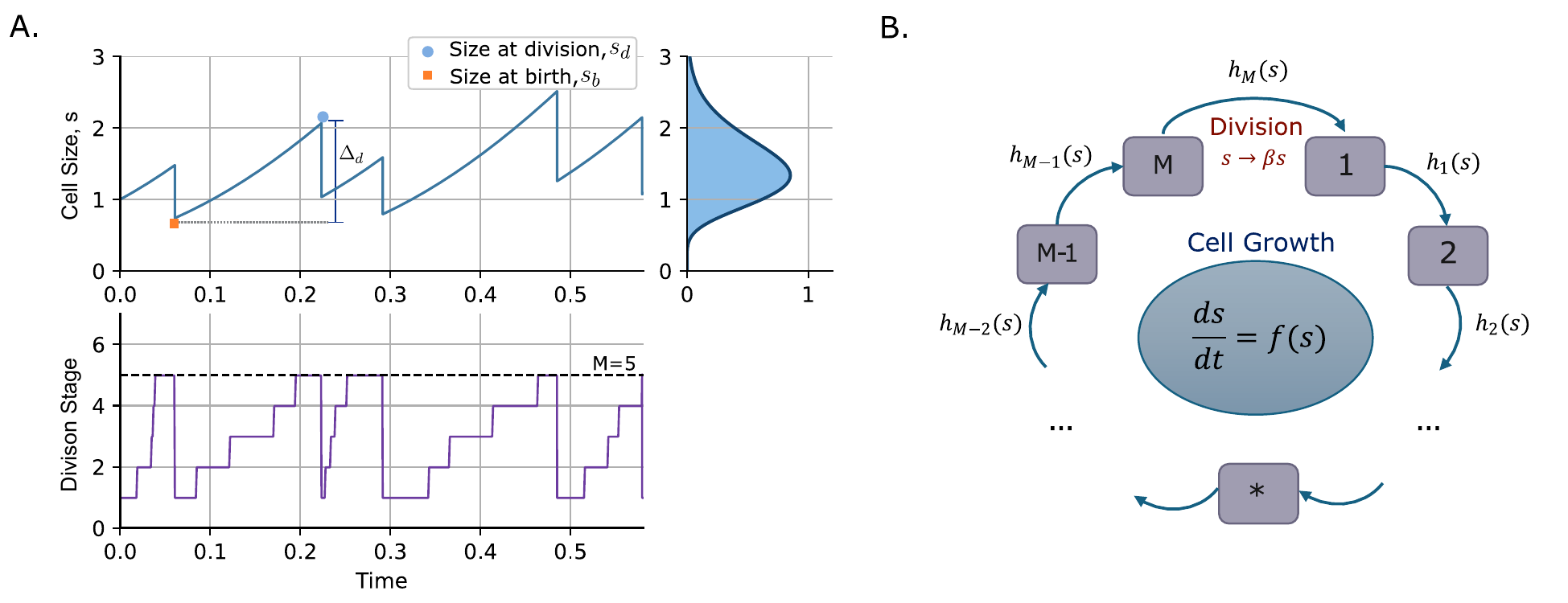}
\caption{\textbf{Cell growth and division cycles captured using the modeling framework of SHS for a multi-step division system.} \textbf{(A)} Example trajectories of cell size (top) and division steps (bottom) along time. Cell size grow at the Hill-type rate given in \eqref{eq:grrate}. In this example division occurs after reaching the 5th stage. The jumps between cycle stages occur randomly at a Hill-type size-dependent rate defined in \eqref{eq:divgrate}. Other cell cycle variables are shown as: size at birth $s_b$, size at division $s_d$, and added size at division $\Delta_d$. \textbf{(B)}
 Schematic of multi-step SHS model describing the cell size dynamics. Size grows continuously based on the given ODE. 
 Division is triggered upon reaching the final stage 
$M$
, following a sequence of stochastic transitions between intermediate stages 
$i\rightarrow i+1$
, each occurring at a size-dependent rate $h_i(s)$ defined in \eqref{eq:divgrate}. Upon division, the cell size undergoes a discrete jump, modeled as $s\rightarrow \beta s$, where $\beta$ is a random variable with mean 
$\langle \beta \rangle$.
}
	\label{fig:Model}
\end{figure*}

We extend the model by representing the cell cycle as a progression through $M$ discrete steps until reaching division. Let us assume a newborn cell starts at stage $i=1$ and transitions sequentially from stage $i$ to $i+1$, when $i \in \{ 1,2, \dots, M-1\}$ at rate $h_i(s)$. Each of these transitions is considered as one step. We define this rate as:
\begin{equation} \label{eq:divgrate}
h_i(s)=\dfrac{k_i g_is}{1+\alpha s}.
\end{equation}
Here, we defined a Bernoulli random process $g_i$, which takes the value $1$ if the cell is in stage $i$ at time $t$, and $0$ otherwise. Only cells in stage $M$ can divide at rate $h_M(s)$. During that division, the cell stage is reset $M\rightarrow1$. Fig. \ref{fig:Model} represents the schematic of explained multi-step SHS model describing the cell growth and division.

The dynamics of $g_i$ described above can be formalized by describing the stage transition following reset maps from stage $i$ to stage $i+1$:
\begin{equation}
    g_i \rightarrow g_i-1, \;\;\;\; g_{i+1}\rightarrow g_{i+1}+1, \;\;\; i \in \{ 1,2, \dots, M-1\}.
\end{equation}
In addition, for the final step, at which the division occurs, the resets are defined by:
\begin{equation} 
    g_M \rightarrow g_M-1, \;\; g_{1} \rightarrow g_{1}+1, \; s \rightarrow \beta s. 
\end{equation}
Note that $\sum_{i=1}^{M} g_i=1$ and the expectation product of $\left\langle g_i g_j \right\rangle =0$ for $i \ne j$. 

Next, we will solve the system in steady-state to obtain its statistical moments. For more clarity and to facilitate closed-form analytical results, we restrict our analysis in this work to the case where all transition rates are identical; i.e., $k_i = k$. 

\section{Quantifying statistical moments in multi-step division model}

\subsection{Steady-state Mean}
Our goal is to determine the mean cell size at steady state.
To do this, we can use the following equation:
\begin{equation}\label{eq:meandef}
     \overline{ \left\langle s \right\rangle } = \sum_{i=1}^{M} \overline{ \left\langle g_i s \right\rangle} ,
\end{equation}
which follows from properties of the Bernoulli random variable $g_i$.
 To find the moments in the right hand-side of~\eqref{eq:meandef}, we start by deriving the moment dynamics for the functions $\log s$, and evaluating it in steady state.
\begin{equation}
    \dfrac{d\left\langle \log s \right\rangle}{dt}=\mu \left\langle \dfrac{1}{1+\alpha s} \right\rangle + k\left\langle \log\beta \right\rangle \left\langle \dfrac{g_M s}{1+\alpha s} \right\rangle, \nonumber
\end{equation}

\begin{equation} \label{eq:gmsa}
     \overline{ \left\langle \dfrac{g_M s}{1+\alpha s} \right\rangle } = \dfrac{-\mu}{k \left\langle \log \beta \right\rangle} \overline{ \left\langle \dfrac{1}{1+\alpha s} \right\rangle }.
\end{equation}
To compute $\left\langle g_M s \right\rangle$, we note that the moment dynamics above yield fractional expressions and therefore, we employ an algebraic manipulation similar to \eqref{eq:trick}. Specifically, this expectation can be rewritten using the following identity:
\begin{small}
\begin{equation}\label{eq:gms1}
    \overline{ \left\langle g_M s \right\rangle } = \overline{\left\langle \dfrac{g_M s}{1+\alpha s} (1+\alpha s) \right\rangle}  = \overline{ \left\langle \dfrac{ g_M s}{1+\alpha s} \right\rangle } +\alpha \overline{ \left\langle \dfrac{ g_M s^2}{1+\alpha s} \right\rangle }.
\end{equation}
\end{small}

Next, to find the higher-order moment showed up in \eqref{eq:gms1}, we write the moment dynamics of $s$ and solve it in steady state:
\begin{equation}
    \dfrac{d\left\langle s \right\rangle}{dt}=\mu \left\langle \dfrac{s}{1+\alpha s} \right\rangle+k (\left\langle \beta \right\rangle-1) \left\langle \dfrac{g_M s^2}{1+\alpha s} \right\rangle,
\end{equation}

\begin{equation}\label{eq:gms2a}
    \overline{ \left\langle \dfrac{g_M s^2}{1+\alpha s} \right\rangle } = \dfrac{-\mu}{k (\left\langle \beta \right\rangle-1)} \overline{ \left\langle \dfrac{s}{1+\alpha s} \right\rangle }. 
\end{equation}
Hence, we can derive a closed-form~expression for  $\left\langle \overline{g_M s}\right\rangle$ by substituting~\eqref{eq:gmsa} and \eqref{eq:gms2a} in \eqref{eq:gms1}.

Our approach is computing all joint moments $\left\langle \overline{g_i s}\right\rangle$ to finally use them in \eqref{eq:meandef}. Thus, we will find the $\left\langle \overline{g_i s}\right\rangle $~for~$i\in\{ 1,2, \dots, M-1\}$.
Similar to \eqref{eq:gms1}, we write:
\begin{small}
\begin{equation}
    \overline{ \left\langle g_i s \right\rangle} = \left\langle \dfrac{g_i s}{1+\alpha s} (1+\alpha s) \right\rangle  = \overline{ \left\langle \dfrac{ g_i s}{1+\alpha s} \right\rangle }+ \alpha \overline{ \left\langle \dfrac{ g_i s^2}{1+\alpha s} \right\rangle}.
\end{equation}
\end{small}

\noindent
The right-hand side terms of the above equation can be generated systematically by writing down the moment dynamics of auxiliary functions in the model. We examine the dynamics of $g_1$ and $g_{i+1}$ for~$i\in\{ 1,2, \dots, M-2\}$:
\begin{equation}
    \dfrac{d\left\langle g_1 \right\rangle}{dt}= -k \left\langle \dfrac{g_1 s}{1+\alpha s} \right\rangle + k \left\langle \dfrac{g_M s}{1+\alpha s} \right\rangle,
\end{equation}
\begin{equation}
    \dfrac{d\left\langle g_{i+1} \right\rangle}{dt}= -k \left\langle \dfrac{g_{i+1} s}{1+\alpha s} \right\rangle + k \left\langle \dfrac{g_i s}{1+\alpha s} \right\rangle.
\end{equation}
Solving the above equations sequentially in steady state gives:
\begin{equation}\label{eq:gms}
     \overline{ \left\langle \dfrac{g_i s}{1+\alpha s} \right\rangle }=  \overline{ \left\langle \dfrac{g_M s}{1+\alpha s} \right\rangle },
\end{equation}
for~$i\in\{ 1,2, \dots, M-1\}$. Additionally, writing the moment dynamics of $\left\langle g_1 s \right\rangle$:
\begin{small}
\begin{equation}
    \dfrac{d\left\langle g_1 s \right\rangle}{dt}=\mu \left\langle \dfrac{ g_1 s}{1+\alpha s} \right\rangle - k \left\langle \dfrac{ g_1 s^2}{1+\alpha s} \right\rangle +k \left\langle \beta \right\rangle \left\langle \dfrac{g_M s^2}{1+\alpha s} \right\rangle,
\end{equation}
\end{small}
and solving it in steady state, we find the following  results:
\begin{small}
\begin{equation}
    \overline{ \left\langle \dfrac{ g_1 s^2}{1+\alpha s} \right\rangle } = \dfrac{\mu}{k} \overline{ \left\langle \dfrac{ g_1 s}{1+\alpha s} \right\rangle }+ \left\langle \beta \right\rangle \overline{ \left\langle \dfrac{ g_M s^2}{1+\alpha s} \right\rangle },
\end{equation}
\end{small}

\begin{equation}
    \overline{ \left\langle \dfrac{ g_{i+1} s^2}{1+\alpha s} \right\rangle } = \dfrac{\mu}{k} \overline{ \left\langle \dfrac{ g_{i+1} s}{1+\alpha s} \right\rangle }+  \overline{ \left\langle \dfrac{ g_i s^2}{1+\alpha s} \right\rangle }.
\end{equation}
\begin{figure}[t]
	\centering
\includegraphics[width=0.9
\linewidth]{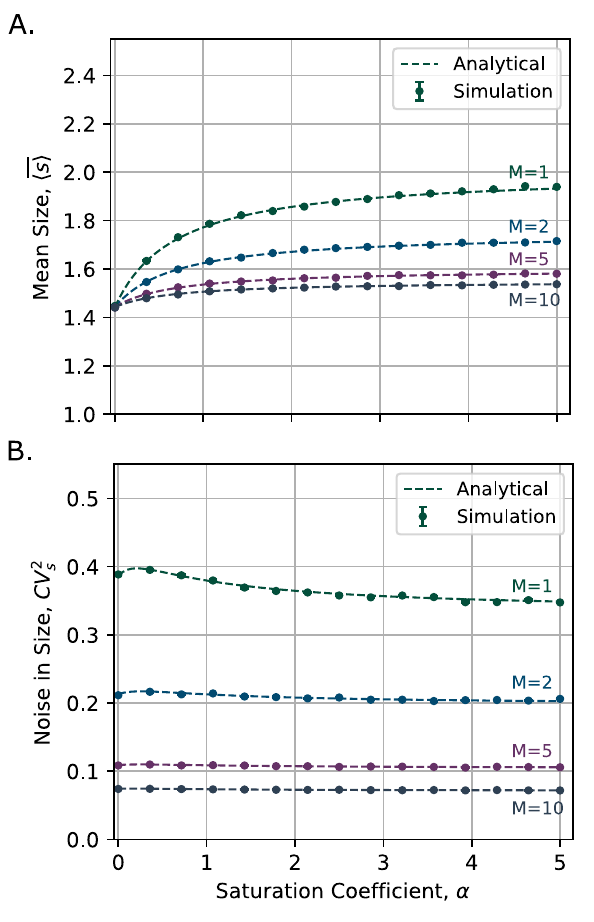}
\caption{ \textbf{Steady-state cell size mean and noise for different number of cycle steps $\boldsymbol{M \in [1,2,5,10]}$ as a function of saturation coefficient~$\alpha$.} \textbf{A)}~Steady-state mean cell size increases with the saturation coefficient and approaches a plateau. The mean size is lower when the number of cycle steps is larger. Analytical predictions from Eq.~\eqref{eq:meanS} (dashed lines) matches with simulation results (dots). \textbf{B)} Steady-state noise in cell size, quantified by coefficient of variation, $CV^2_s$ describd in \eqref{eq:cv}, rises slightly with the saturation coefficient, exhibits a small peak, and then levels off. Noise decreases as the number of cycle steps increases. Results are based on simulations of 2000 cells with parameters as $\mu=\log 2$, $k=\mu M$ and $ \beta=1/2$ with probability 1.}
	\label{fig:MeanCV}
\end{figure}

\noindent
Finally, substituting the results in \eqref{eq:meandef} and simplifying, we obtain:
\begin{equation} \label{eq:giss}
    \overline{ \left\langle g_i s \right\rangle }=\left(1+i \alpha \dfrac{\mu}{k}\right) \overline{ \left\langle \dfrac{g_M s}{1+\alpha s} \right\rangle }+ \alpha \langle \beta \rangle \overline{ \left\langle \dfrac{g_M s^2}{1+\alpha s} \right\rangle },
\end{equation}
for~$i\in\{ 1,2, \dots, M-1\}$.
\noindent
The final step is to calculate the unknown moments of $ \scriptstyle  \overline{ \left\langle \dfrac{1}{1+\alpha s} \right\rangle }$ and $\scriptstyle \overline{ \left\langle \dfrac{s}{1+\alpha s} \right\rangle}$ which appeared in equations \eqref{eq:gmsa} and \eqref{eq:gms2a}. These terms represent higher-order correlations between cell size and growth-rate saturation and are not directly accessible from the lower-order moments. To close the system, we exploit algebraic and probabilistic relations inherent to the stochastic hybrid system, which allow these expectations to be expressed exactly in terms of known model parameters and step-dependent moments. Specifically, we note that:
\begin{equation}\label{eq:moms2}
     \overline{ \left\langle \dfrac{s}{1+\alpha s} \right\rangle }= \sum_{i=1}^{M}{\overline{ \left\langle \dfrac{g_i s}{1+\alpha s} \right\rangle}}.
\end{equation}
Using expressions \eqref{eq:moms2}, \eqref{eq:gms} and the identity \eqref{eq:trick} for $n=1$, we obtain:

\begin{equation}
    \overline{  \left\langle \dfrac{1}{1+\alpha s} \right\rangle } = \dfrac{k \left\langle \log \beta \right\rangle }{k \left\langle \log \beta \right\rangle -M \alpha \mu}.
\end{equation}
Substituting these relations into \eqref{eq:giss} yields a fully determined system that can be solved explicitly.
Finally, from~\eqref{eq:meandef} we can write: 
\begin{equation}
     \overline{ \left\langle s \right\rangle } = \sum_{i=1}^{M} \overline{ \left\langle  g_i s \right\rangle } = (M-1)\sum_{i=1}^{M-1} \overline{\left\langle  g_i s \right\rangle }+ \overline{ \left\langle  g_M s \right\rangle }.
\end{equation}

This procedure results in a general expression for the mean cell size at steady state, which accounts for the number of steps $M$, the stage transition rate constant $k$, and the parameters governing growth dynamics, $\mu$ and $\alpha$:
\begin{equation} \label{eq:meanS}
    \overline{ \left\langle s \right\rangle } = \dfrac{\mu M \left( -2 k (\left\langle\beta \right\rangle-1) + \alpha \mu \left( 1+ \left\langle \beta \right\rangle (M-1)+M \right) \right)}{2 k (\left\langle \beta \right\rangle-1) \left(k \left\langle \log \beta \right\rangle - M \alpha \mu \right)}.
\end{equation}
Fig. \ref{fig:MeanCV}A shows  the steady-state mean cell size, comparing the analytical expression from Eq.~\eqref{eq:meanS} with simulation results.
As the saturation coefficient increases, the mean cell size rises and eventually approaches a plateau. Additionally, the mean size decreases with an increasing number of~cycle~steps.

\subsection{Steady-state Cell Size Noise}
In this section, we aim to quantify the noise in cell size, defined as the variability observed along single-cell lineages. To achieve this, we extend the previously discussed framework to higher-order statistics. Specifically, following the same approach as in the preceding section, we derive the moment dynamics for 
 $s^2$  and related functions, which enables us to compute the second-order moment  $\left\langle s^2 \right\rangle$ using the following equation:
\begin{equation}
     \overline{ \left\langle s^2 \right\rangle } = \sum_{i=1}^{M} \overline{ \left\langle  g_i s^2 \right\rangle  } = (M-1)\overline{ \left\langle  g_i s^2 \right\rangle }+ \overline{ \left\langle  g_M s^2 \right\rangle }.
\end{equation}

\noindent
Due to space limit and avoiding repetition, here we represent the final result we obtained for the steady-state second-order moment:
{
\begin{align}
\overline{ \left\langle s^2 \right\rangle }=
&\frac{
\mu 
}{
6 k^2 (\langle \beta^2 \rangle-1)  (k \log \beta - \alpha \mu M)
} \nonumber \\ \nonumber
&[
\mu M (
-3 (\langle \beta \rangle -1) k (-3 + 3 \langle \beta \rangle (M-1) + M)\\ \nonumber
&+ 2 \alpha \mu (\langle\beta \rangle+1) ( M-1) (1 + M + \langle\beta \rangle(2 M-1))
) \\
&+ 12 k (1 + \langle \beta^2 \rangle ( M-1)) (-k \log \beta + \alpha \mu M) \, \overline{\left\langle s\right\rangle}
].
\end{align}
}

\noindent
We illustrated in Fig. \ref{fig:MeanCV}B the steady-state noise in cell size, $CV^2_s$, as described in Eq.~\eqref{eq:cv}, while comparing it with the simulation results. We observe that cell size noise decreases as the number of cycle
steps increases. Since the analytical expressions are derived exactly without approximation, they show a perfect match with the simulation~data.

\begin{figure*}[t!] 
	\centering
\includegraphics[width=0.8
\linewidth]{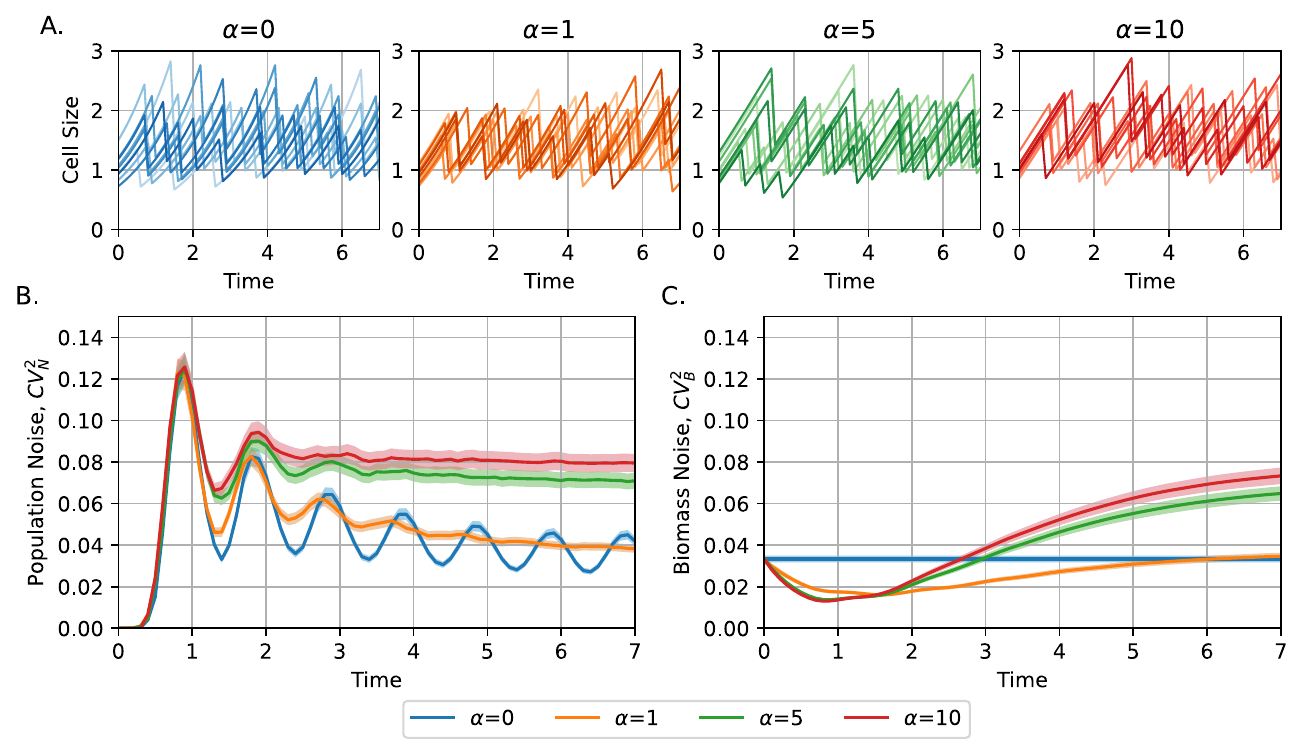}
\caption{\textbf{Increasing the saturation coefficient in the growth and division rates leads to a higher steady-state noise in population. } \textbf{A)} Examples of cell size trajectories for different saturation coefficients, $\alpha \in {[0,1,5,10]}$ used in growth law \eqref{eq:grrate} and division rate \eqref{eq:divgrate}. The case of $\alpha=0$ corresponds to exponential growth, while larger $\alpha$ values approximate linear growth. \textbf{B)} Time dynamics of  population noise , quantified by $CV^2_N$, for varying values of $\alpha$.  Exponential growth ($\alpha=0$) yields the lowest steady-state noise level. Population is defined as the number of individuals in a colony. \textbf{C)} Time dynamics of biomass noise, quantified by $CV^2_B$, for varying values of $\alpha$. Biomass is defined as the sum of all cell size in individuals and the noise is calculated over different colonies at a given time. Simulations are performed assuming $M=10$ steps, $k=\mu M$, and $\mu$ is chosen such that the population doubles once per unit time. Statistics estimated using 10000 colonies with random progenitor cell size with statistics $\langle s\rangle$=1 and $CV^2_{s}=\frac{1}{3M}$ as found in \cite{modi2017analysis}. Cells divide in half deterministically.}
	\label{fig:pop}
\end{figure*}

\section{Effect of saturation on population-level noise}

Building on our previous analysis of cell size statistics along single lineages in a multi-step system, we now will extend the study to the population level. Specifically, we will investigate how saturation in growth and division rates influences fluctuations across entire populations composed of multiple colonies.
Each colony is defined as the set of descendants originating from a single ancestor cell born at time zero, with an initial size sampled from the distribution of newborn cells and in  stage $i=1$. At each time point, we compute the mean and noise of individuals $N$ across all colonies to characterize population-level~variability.

Fig. \ref{fig:pop} 
represents simulation results illustrating the time dynamics of noise in population, quantified using the squared coefficient of variation of individuals in the colony ($CV^2_N$), for varying values of the saturation coefficient $\alpha$.
Our results reveal that increasing the saturation coefficient in growth and division rates leads to a higher steady-state noise in the population. Additionally, higher values of  $\alpha$ reduce the settling time of oscillatory transients. Among the growth laws considered, exponential growth ($\alpha=0$) yields the lowest steady-state levels of population noise.

We also illustrated the cell size noise in biomass $CV^2_B$.
Biomass $B$ is defined as the total sum of cell sizes in a colony, averaged across replicates (see a detailed explanation in \cite{nieto2024generalized}). For exponentially growing cells, biomass increases deterministically and exponentially. Therefore, the biomass noise is constant over time and equal to the initial biomass noise, that is, the cell size noise of the newborn progenitors. In contrast, under non-exponential growth laws $\alpha>0$, biomass behaves as a stochastic process due to variability in individual cell trajectories (the total biomass growth is not exponential, but depends on the size of each descendant).
In particular, as explained previously \cite{nieto2024generalized}, we observe that the noise of the population-individuals converges to the noise of biomass when $t \rightarrow \infty$ for all growth laws. This analysis of the proliferation dynamics under our proposed saturated growth laws reveals how cells can exhibit different proliferation dynamics for particular growth laws, even though they follow a similar division strategy (adder).

\section{Conclusions}
In this study, we developed a stochastic multi-step model for cell size control, extending the classical adder framework to account for general growth laws with saturation. By implementing a size-dependent Hill-type growth law, we capture the biologically relevant scenario in which very large cells exhibit linear-like growth, introducing a saturation effect that modulates size dynamics.
Additionally, the multi-step formulation allows us to incorporate the sequential nature of division, reflecting the reality that cells progress through internal stages before dividing. This approach yields exact analytical expressions for cell size moments, providing a quantitative understanding of how growth saturation influences both mean size and size fluctuations (noise). We find that stronger growth saturation increases the mean cell size while slightly reducing variability compared to the classical exponential growth scenario. Notably, the adder property of a constant added size is preserved, highlighting that these changes in noise are a direct consequence of the growth law rather than trivial scaling with the mean. Together, these results show that the stochastic multi-step adder framework generalizes classical adder theory to accommodate size homeostasis across diverse~growth regimes.

By linking single-cell dynamics to stochastic clonal proliferation, our results further demonstrate that growth saturation shapes population-level variability. Colonies derived from individual cells under higher saturation conditions show higher fluctuations, emphasizing that single-cell regulatory mechanisms propagate to affect clonal heterogeneity.
These insights bridge mechanistic models of cell size control with experimentally observed variability at the population level, enabling testable predictions across diverse organisms. Future work will focus on explicitly modeling population dynamics to compare fluctuations in population size with experimental data, as well as investigating the impact of population-level randomness in fluctuation assays, such as experiments probing the heritability of gene expression
\cite{kempe2015volumes, weidemann2023minimal, claude2021transcription, padovan2015single}.
\section{Acknowledgments}
This work is supported by NIH-NIGMS via grant R35GM148351.

\bibliographystyle{ieeetr}
\bibliography{Ref} 

\end{document}